# Free particle trapped in an infinite quantum well examined through the discrete calculus model

Dušan POPOV

**Politehnica University Timisoara, Department of Physical Foundations of Engineering, Bd. V. Parvan No. 2, 300223 Timisoara, Romania**
**E-mail: dusan_popov@yahoo.co.uk**
**ORCID:** https://orcid.org/0000-0003-3631-3247.

**Abstract** In the present paper we use the discrete approach in order to solve the Schrödinger as well as the Bloch equations for a free particle and the „quantum gas" of free particles embeded in an infinite quantum well with the finite width, respectively. We obtain the expressions of energy eigenvalues, the eigenfunctions as well as the density matrix and partition function for the discrete case. By applying the so called continuous quantum mechanics limit we recover the corresponding results in the continuous-variable quantum mechanics.

1. **Introduction**

The discrete quantum mechanics (dQM) is the counterpart of the ordinary or continuous-variable quantum mechanics (cQM) in which the second order differential time-independent Schrödinger equation is written as a second order difference equation. The dQM was developed by Odake and Sasaki [Odake, 2011]. In this context the dQM can be considered as a generalization (or deformation) of the ordinary or continuous-variable QM, in the sense that a differential equation can be obtained from a difference equation by applying an appropriate limit. In what follows, we will call this limit the *continuous quantum mechanical limit* ($\lim_{cQM}$). Generally, by applying this limit to the results obtained in dQM, you need to get the appropriate results in cQM.

On the one hand, there are only a few physical models for which the Schrödinger equation, even in one dimension, can be solved analytically in cQM, that is, they admit precise analytic solutions. Consequently, in these cases the most interesting physical problems must be solved numerically, considering the unprecedented development of computers and their programming languages.

On the other hand, intrinsic structure of matter has a discontinuous character, so an increasing use of discrete models is absolutely desirable. Moreover, the discrete approach is suitable for many different applications in solid state physics, since many physical systems are composed of crystalline semiconductors, where the distances between atoms or ions are finite. Also, the discrete approach can be successfully used in various nanostructures and nano-objects including quantum wells, wires and dots [Hilder, 2007].

The cQM can be considered, from one point of view, as an average behavior, and at the limit of a very large number of particles the results from dQM approach those from cQM [Baowan, 2016].

The purpose of the present paper is to show that a discrete finite difference calculus can be successfully applied also to some (simple) quantum mechanical models. The obtained results are, evident, different comparing with there obtained with continuous quantum mechanical methods. But, using the continuous quantum mechanical limit, all the obtained results in the discrete approach tend, to the corrresponding results of the continuous space models.

In the present paper we have chosen a simple physical model: a free particle trapped in an infinite square quantum well. Even though the quantum dynamics in an infinite well potential represents a rather unphysical limit situation, it is a familiar textbook problem and a simple tractable model for the behavior of a quantum particle. The infinite square quantum potential well model is not totally applied to any real situation, but it will be in a certain sense a good approximation to some other simple quantum models. Actually, the same role is played by the harmonic oscillator model: the behavior of real quantum systems (e.g. molecular vibrations) is harmonic only near the equilibrium position, but still the harmonic oscillator model is used successfully.

## 2. Preliminaries related to the discrete operations

In order to apply the dQM for some physical models, it will be useful to remember some rules of the discrete calculus and the difference equations (see, [Elaydi, 2005], [Agarwal, 2000] and the references therein and also [Jensen, 2011], [Tošić, 2007]). As it is well-known, the ordinary (continuous) derivative a function $f(x)$ which is analytics in the neighborhood of a point $x$, with respect to the continuous variable $x$ is defined as

$$\frac{df(x)}{dx} = \lim_{\Delta x \to 0} \frac{\Delta f(x)}{\Delta x}, \qquad (2.1)$$

where

$$\frac{\Delta f(x)}{\Delta x} = \frac{f(x+\Delta x) - f(x)}{\Delta x}. \qquad (2.2)$$

Hovewer, the computers cannot deal with the infinitesimal limit $\Delta x \to 0$ and, consequently, it is necessary to define the discrete counterparts of the continuous derivative. So, if the variable $x$ is discrete, i.e. the functions values are available on a discrete set of points, from the map $x \to x_n = an$, where $\Delta x \equiv a$ is a positive constant (the so called lattice spacing), then for a function $f(x_n)$ which is analytic in the neighborhood of a point $x_n$ it can define their discrete counterpart of the step $a$ as

4$$\frac{\Delta}{\Delta x_n} f(x_n) = \frac{f(x_{n+1}) - f(x_n)}{x_{n+1} - x_n} \quad \rightarrow \quad \frac{1}{a}\frac{\Delta}{\Delta n} f(n) = \frac{f(n+1) - f(n)}{a}. \tag{2.3}$$

Practically, they can define three discrete valid counterparts of the continuous derivatives: the forward discrete derivative (FDD), the backward discrete derivative (BDD), and also the central or centered discrete derivative (CDD), since it uses forward, backward or central differencing [Elaydi, 2005], [Agarwal, 2000]. All these three definitions are equivalent in the continuous case, but lead to *different aproximations* in the discrete case.

For the purpose of our paper it is useful to use the *centered discrete derivative* defined with the step $2a$ as follows:

$$\frac{\Delta}{\Delta x_n} f(x_n) = \frac{f(x_{n+1}) - f(x_{n-1})}{x_{n+1} - x_{n-1}} \quad \rightarrow \quad \frac{1}{a}\frac{\Delta}{\Delta n} f(n) = \frac{f(n+1) - f(n-1)}{2a}. \tag{2.4}$$

The motivation of the use of centered discrete derivative consists in the fact that, using this kind of derivative, the solutions of the difference Schrödinger equation for the examined case have an oscillatory character [Jannussis, 1980].

In order to simplify the notations and not to create confusion, we will use the following notation for the discrete forward derivative

$$\frac{\Delta}{\Delta x_n} f(x_n) \equiv \frac{1}{a}\frac{\Delta}{\Delta n} f(n) = \frac{f(n+1) - f(n-1)}{2a}. \tag{2.5}$$

Consequently, the second central or centered derivative (in what follows we will omit the words „central" or „centered") is

$$\left(\frac{\Delta}{\Delta x_n}\right)^2 f(n) \equiv \left(\frac{1}{a}\frac{\Delta}{\Delta n}\right)^2 f(n) = \frac{f(n+2) - 2f(n) + f(n-2)}{4a^2}. \tag{2.6}$$

Because we have to deal with physical observables, it should be paid attention to the dimensions of the physical observables which participate to the difference equations (so, it must appear, to the right place, the latice spacing $a$).

In another train of thoughts, a translation operator with the step $a$, i.e. $T_a \equiv T_1$ act on an analytical function as follows:

$$T_1 f(x_n) = f(x_{n+1}) \quad \rightarrow \quad T_1 f(n) = f(n+1) \tag{2.7}$$

and it has the following properties [Jensen, 2011], [Tošić, 2007]:

$$(T_1)^{-1} = T_{-1} \; ; \quad T_1 T_{-1} = T_{-1} T_1 = 1 \; ; \quad (T_m)^n = T_{m \cdot n} \; , \quad m, n \in \mathbb{N}^* . \tag{2.8}$$

Consequently, the first and second difference derivatives can be written with the help of the translation operator:

$$\frac{1}{a} \frac{\Delta}{\Delta n} f(n) = \frac{1}{2a} (T_1 - T_{-1}) f(n) \; ; \quad \left( \frac{1}{a} \frac{\Delta}{\Delta n} \right)^2 f(n) = \frac{1}{4a^2} (T_1 - T_{-1})^2 f(n) . \tag{2.9}$$

As is known, the integral of a continuous function is

$$\int dx \, f(x) = F(x) + C \tag{2.10}$$

where the function $F(x)$ is the antiderivative function of $f(x)$, i.e.:

$$\frac{dF(x)}{dx} = f(x) . \tag{2.11}$$

or, for discrete variable

$$\frac{1}{a} \frac{\Delta}{\Delta n} F(n) = f(n) . \tag{2.12}$$

For the discrete variable $x_n = an$, instead of the integration symbol $\int dx \ldots$ (which is the inverse of the differentiation symbol $\frac{d}{dx}$), which, formally, can be written as $\int dx \ldots = \left( \frac{d}{dx} \right)^{-1} \ldots$ , we will use the *symbol* $\left( \frac{\Delta}{\Delta x_n} \right)^{-1}$, as the inverse of the discrete differentiation symbol $\frac{\Delta}{\Delta x_n} \equiv \frac{1}{a} \frac{\Delta}{\Delta n}$ where $\left( \frac{\Delta}{\Delta x_n} \right)^{-1} \equiv \left( \frac{1}{a} \frac{\Delta}{\Delta n} \right)^{-1}$ is *only the discrete dimensionless integration symbol* (equivalent to the graphic symbol of integration $\int$ ) which evinces that the discrete integrals refer to the variable $x_n = na$.

Formally, we have

$$\frac{\Delta}{\Delta x_n} \left( \frac{\Delta}{\Delta x_n} \right)^{-1} = \frac{1}{a} \frac{\Delta}{\Delta n} \left( \frac{1}{a} \frac{\Delta}{\Delta n} \right)^{-1} = 1 \tag{2.13}$$

For the definite integration, obviously we have to attach the integration limits. So, it means that:

$$\left(\frac{\Delta}{\Delta x_n}\right)^{-1}\bigg|_{x_{min}}^{x_{max}} f(x_n) = F(x_{max}) - F(x_{min}) \qquad (2.14)$$

and also

$$\left(\frac{1}{a}\frac{\Delta}{\Delta n}\right)^{-1}\bigg|_{n_{min}}^{n_{max}} f(n) = F(n_{max}) - F(n_{min}) \qquad (2.15)$$

One way to define the integration is to find the inverse transform of the derivative and this concept is known as finding the antiderivative. If we apply to the above relation to the inverse derivative operator (the integration operator) we recover the antiderivative function:

$$\left(\frac{\Delta}{\Delta x_n}\right)^{-1} \frac{\Delta}{\Delta x_n} F(x_n) = F(x_n) = \left(\frac{\Delta}{\Delta x_n}\right)^{-1} f(x_n). \qquad (2.16)$$

$$\left(\frac{1}{a}\frac{\Delta}{\Delta n}\right)^{-1} \frac{1}{a}\frac{\Delta}{\Delta n} F(n) = F(n) = \left(\frac{1}{a}\frac{\Delta}{\Delta n}\right)^{-1} f(n). \qquad (2.17)$$

Applying to Eq. (2.16) the derivative operator, we successively obtain:

$$\frac{\Delta}{\Delta x_n}\left(\frac{\Delta}{\Delta x_n}\right)^{-1} f(x_n) = f(x_n) = \frac{\Delta}{\Delta x_n} F(x_n) = \frac{F(x_{n+1}) - F(x_{n-1})}{2\Delta x} = \frac{(T_1 - T_{-1})F(x_n)}{2\Delta x}, \qquad (2.18)$$

from which, applying the inverse difference operator $\left(\frac{\Delta}{\Delta x_n}\right)^{-1}$, it results

$$F(x_n) = 2\Delta x (T_1 - T_{-1})^{-1} f(x_n) \quad , \qquad F(n) = 2a(T_1 - T_{-1})^{-1} f(a) \qquad (2.19)$$

Finally, discrete integration is performed using the following relation:

$$\left(\frac{\Delta}{\Delta x_n}\right)^{-1} f(x_n) = 2\Delta x (T_1 - T_{-1})^{-1} f(x_n) \qquad (2.20)$$

If we work with physical quantities, we must take into account their dimensions:

$$\left(\frac{1}{a}\frac{\Delta}{\Delta n}\right)^{-1} f(n)\bigg|_{n_{min}}^{n_{max}} = 2a(T_1 - T_{-1})^{-1} f(n)\bigg|_{n_{min}}^{n_{max}} \qquad (2.21)$$

In order to be able to calculate the right hand side of the above relation, we must expand the operator expression in a power series of the translation operator $T_1$, using their above mentioned properties:

$$(T_1 - T_{-1})^{-1} = -\frac{1}{T_{-1}(1-T_2)} = -\frac{1}{T_{-1}}\sum_{k=0}^{\infty} T_{2k} = -\sum_{k=0}^{\infty} T_{2k+1}. \quad (2.22)$$

Supposing that the series is convergent, the discrete integral can be calculated by using the following equation:

$$\left(\frac{1}{a}\frac{\Delta}{\Delta n}\right)^{-1} f(n)\bigg|_{n_{\min}}^{n_{\max}} = -2a \sum_{k=0}^{\infty} T_{2k+1} f(n)\bigg|_{n_{\min}}^{n_{\max}} \quad (2.23)$$

Let us provide some examples of using the above formula by indicating the results of several discrete integrals that will be useful in the following sections, whose validity can be directly proved by performing the discrete differentiation operations (for brevity we choose $a=1$ in front). So, the results are:

$$\left(\frac{\Delta}{\Delta n}\right)^{-1} 1 = n, \quad \left(\frac{\Delta}{\Delta n}\right)^{-1} \cos n\alpha = \frac{\sin n\alpha}{\sin \alpha}, \quad \left(\frac{\Delta}{\Delta n}\right)^{-1} \sin n\alpha = -\frac{\cos n\alpha}{\sin \alpha} \quad (2.24)$$

$$\left(\frac{\Delta}{\Delta n}\right)^{-1} \sin^2 n\alpha = \frac{n}{2} - \frac{1}{2}\frac{\sin 2n\alpha}{\sin 2\alpha}. \quad (2.25)$$

With these mathematical preliminaries we are able to attack the simple quantum-mechanical models which are based upon the resolution of one-dimensional time-independent Schrödinger equation.

### 3. Remembering the notions of the discrete Schrödinger equation

Let us now remember the main elements (known before, see, [Jannussis, 1980], [Boykin, 2004], [Wolf, 1990]) of the finite difference calculation applied to solving the Schrödinger equation for the infinite square quantum well. We will then use the obtained results to solve the finite difference Bloch equation, which is the main purpose of the paper.

A simple one-dimensional time-independent differential Schrödinger equation for a particle with the effective mass $m^*$ in a potential field $U(x)$, where $x$ is a continuous variable, i.e.:

$$-\frac{\hbar^2}{2m^*}\frac{d^2}{dx^2}\Psi(x) + U(x)\Psi(x) = E\Psi(x) \tag{3.1}$$

can be written as a difference equation with the following structure

$$-\frac{\hbar^2}{2m^*}\frac{\Delta^2}{\Delta x_n^2}\Psi(x_n) + U(x_n)\Psi(x_n) = E\Psi(x_n) \tag{3.2}$$

or, more convenient, through the *central derivative* of second order with step $2a$:

$$\frac{1}{4}[\Psi(n+2) - 2\Psi(n) + \Psi(n-2)] + \frac{2m^*}{\hbar^2}a^2[E - U(n)]\Psi(n) = 0. \tag{3.3}$$

In order to work with a dimensionless equation, we can perform the following notations to use the dimensionless quantities:

$$\frac{2m^*E}{\hbar^2}a^2 = \tilde{E}, \qquad \frac{2m^*}{\hbar^2}a^2 U(n) = \tilde{U}(n) \tag{3.4}$$

Consequently, the discrete Schrödinger equation can be written as

$$\frac{1}{4}[\Psi(n+2) - 2\Psi(n) + \Psi(n-2)] + [\tilde{E} - \tilde{U}(n)]\Psi(n) = 0. \tag{3.5}$$

Our task is to apply the discrete version of solving the Schrödinger equation to an analytically solvable quantum-mechanical problem – *the infinite square well*.

Let us we consider a free quantum particle of the effective mass $m^*$ moving in an asymetric infinitely deep square well potential, i.e. trapped in a spatial region of width $L = N a = \text{const.}$, where $a = x_{n+1} - x_n$, $n = 0, 1, 2, ..., N$, which is defined as follows:

$$U(x_n) = \begin{cases} 0, & \text{if } 0 \leq x_n \leq L = N a \\ +\infty, & \text{otherwise} \end{cases} \tag{3.6}$$

Inside of the well, in which $U = 0$, the discrete Schrödinger equation becomes

$$\frac{1}{4}[\Psi(n+2) - 2\Psi(n) + \Psi(n-2)] + \tilde{E}\Psi(n) = 0. \tag{3.7}$$

This is a homogenous difference equation of the second order with the constant coefficients.

The use of the central difference formula for the second derivative of the Hamiltonian is justified by the fact that in this manner we lead to the Hermitian Hamiltonian (as is stated also in Ref. [Boykin, 2004], which refers to the connection with the tight-binding model). This can be seen immediately, if we transcribe the above equation in the matrix form

$$\frac{1}{4}\begin{pmatrix} -2 & 0 & 1 & 0 & \ddots & 0 \\ 0 & -2 & 0 & 1 & \ddots & \ddots \\ 1 & 0 & \ddots & \ddots & \ddots & 0 \\ 0 & 1 & \ddots & -2 & \ddots & 1 \\ \ddots & \ddots & \ddots & \ddots & \ddots & 0 \\ 0 & \ddots & 0 & 1 & 0 & -2 \end{pmatrix} \begin{pmatrix} \Psi(0) \\ \Psi(1) \\ \vdots \\ \Psi(n) \\ \vdots \\ \Psi(N) \end{pmatrix} = -\tilde{E} \begin{pmatrix} \Psi(0) \\ \Psi(1) \\ \vdots \\ \Psi(n) \\ \vdots \\ \Psi(N) \end{pmatrix} \qquad (3.8)$$

where we can observe that this Hamiltonian matrix is three-diagonal, and, of course, Hermitian ($H_{ij} = H_{ji}^*$). This leads to the oscillatory character of solutions [Jannussis, 1980].

Therefore, we will try the general solution of the difference equation as

$$\Psi(n) = C_1 \cos(n\varphi) + C_2 \sin(n\varphi) \ . \qquad (3.9)$$

To find the arbitrary integration constants $C_{1,2}$ we will use the border conditions: at the borders of the infinite square well the eigenfunctions must vanish (particle is confined to a space of the width $L = Na$ surrounded by infinite square potential well). So, we have:

$$\begin{aligned} \Psi(n=0) = 0 &\Rightarrow 0 = C_1 \\ \Psi(n=N) = 0 &\Rightarrow 0 = C_2 \sin N\varphi \end{aligned} . \qquad (3.10)$$

The last condition leads to the energy quantization, because we must have $N\varphi = n_E \pi$, where $n_E = 0, 1, 2, \ldots$ will play the role of the main (principal) quantum number. From this it follows: $\varphi = \pi \frac{n_E}{N}$ and the eigenfunctions are

$$\Psi_{n_E}(n) = C_2 \sin\left(\pi \frac{n_E}{N} n\right) . \qquad (3.11)$$

Introducing this expression in the discrete Schrödinger equation (3.7), we obtain

$$\sin[(N+2)\varphi] - 2\sin(N\varphi) + \sin[(N-2)\varphi] + 4\tilde{E}\sin(N\varphi) = 0 \qquad (3.12)$$

and also, after some trigonometric manipulations and because the coefficient of $\sin(N\varphi)$ must vanish, we must have

$$-2\cos 2\varphi + 2 + 4\tilde{E} = 0. \tag{3.13}$$

This condition leads to the quantized energy eigenvalues of the free particle in the infinite square well:

$$E_{n_E}^{(d)} = \frac{\hbar^2}{2m^*a^2}\sin^2\left(\pi\frac{n_E}{N}\right), \tag{3.14}$$

where, we specify once again, $n_E = 1, 2, 3, \ldots$ play the role of the *principal quantum number*.

We will point out that this result was obtained earlier by Wolf [Wolf, 1990] which used also the central discrete derivative, but with step $a/2$, which confirms that, due to the symmetry in the definition of central discrete derivative, the used step does not play an essential role.

Note that, as well as for the continuous case, for the value $n_E = 0$ the eigenfunction $\Psi_0(n) = 0$ and also the corresponding energy is $E_0^{(d)} = 0$, and consequently, the ground state is obtained for $n_E = 1$. The solution where $n_E = 0$ gives the null wave function (and zero energy), which can be interpreted as no solution. Consequently, $n_E = 1, 2, \ldots, \infty$.

Evidently, this expression is different from their counterpart in the continuous quantum mechanics (cQM) (i.e. there obtained by the resolution of the Schrödinger equation for the continuous variable $x$). If the number $N$ of lattice points becomes infinite, the spacing between them becomes infinitesimal. Imposing that the quantum well width remains constant and finite, we can perform the following limit [Wolf, 1990], [Boykin, 2005]: $N \to \infty, a \to 0, aN = L$ (finite). We call this limit *the continuous quantum mechanical limit* ($\lim_{cQM}$). By applying this limit we can verify the correctness of the obtained results for the discrete case.

For the energy eigenvalues $E_{n_E}$, by using the l'Hospital rule for the indeterminate forms of limits, this limit leads to the correct result for the continuous case:

$$\lim_{\substack{N\to\infty \\ a\to 0 \\ aN=L=\text{const}}} E^{(d)}_{n_E} \equiv \lim_{cQM} E^{(d)}_{n_E} = \frac{\hbar^2}{2m^*} \lim_{cQM} \left[\frac{\sin\left(\frac{n_E}{aN}\pi a\right)}{a}\right]^2 = \frac{\hbar^2}{2m^*}\frac{n_E^2\pi^2}{(aN)^2} = E^{(c)}_{n_E}. \qquad (3.15)$$

In this manner, the expression for the energy eigenvalues of a free particle in a discrete infinite quantum well reduces to the well known parabolic expression for the energy eigenvalues of a free particle in a continuous infinite quantum well (see, e.g. [Dong, 2007] or another quantum mechanics book):

$$E^{(c)}_{n_E} = \frac{\hbar^2\pi^2}{2m^*L^2}n_E^2 . \qquad (3.16)$$

The normalization constant $C_2$ is obtained from the normalization condition of the discrete eigenfunctions $\Psi(n)$:

$$a\left[\frac{\Delta^{-1}}{\Delta n}|\Psi(n)|^2\right]\Bigg|_0^N = 1 . \qquad (3.17)$$

We have, successively:

$$\frac{\Delta^{-1}}{\Delta n}|\Psi(n)|^2 = |C_2|^2 \frac{\Delta^{-1}}{\Delta n}\sin^2\left(\pi\frac{n_E}{N}n\right) = \frac{1}{2}|C_2|^2\left[n - \frac{\sin 2\pi\frac{n_E}{N}n}{\sin 2\pi\frac{n_E}{N}}\right]. \qquad (3.18)$$

Consequently, the definite integral is then:

$$a\left[\frac{\Delta^{-1}}{\Delta n}|\Psi(n)|^2\right]\Bigg|_0^N = \frac{a}{2}|C_2|^2\left[N - \frac{\sin 2\pi\frac{n_E}{N}N}{\sin\pi\frac{n_E}{N}}\right] = \frac{a}{2}|C_2|^2 N = 1, \qquad (3.19)$$

so that we obtain the same value for the normalization constant $C_2$ as in the continuous case:

$$|C_2| = \sqrt{\frac{2}{aN}} = \sqrt{\frac{2}{L}} . \qquad (3.20)$$

Finally, the normalized eigenfunctions for the discrete case are identical to those of the continuous one:

$$\Psi_{n_E}(n) = \sqrt{\frac{2}{L}}\sin\left(\pi\frac{n_E}{N}n\right) , \qquad (3.21)$$

i.e. it is the solution of sinusoidal stationary waves along the $x$- axis, with $0 \leq x_n \leq L$ and $n_E = 1, 2, ..., \infty$. This result coincides with the one in [Tarasov, 2016] (for the initial phase $\phi_0 = -\pi/2$), where the discretization was done by another method.

### 4. The discretized Bloch equation for the infinite square well

Let us consider a „quantum ideal gas" of the free particles trapped in the infinite discrete quantum well which are in the thermodynamical equilibrium with the „reservoir" (of finite width and confined by the infinitely higher walls) at the temperature $T$ (with $T = (\beta k_B)^{-1}$, where $k_B$ is the Boltzmann's constant). Obviously, such a gas is in a mixed quantum state described by the statistical or density operator $\rho$ or, in some representation, say $\xi$, by the density matrix $\rho(\xi, \xi'; \beta)$.

For the quantum system described by the Hamiltonian $H$ and which obeys the canonical distribution, the canonical density operator $\rho$ is an exponential operator

$$\rho = e^{-\beta H} \qquad (4.1)$$

and satisfies the quantum Bloch equation with the initial condition:

$$-\frac{\partial}{\partial \beta}\rho = H\rho, \qquad \rho(\beta = 0) = 1 \quad . \qquad (4.2)$$

The canonical (or unnormalized) density matrix $\rho(\xi, \xi'; \beta)$ can be normalized to unity and the corresponding normalized density matrix $\tilde{\rho}(\xi, \xi'; \beta)$ is

$$\tilde{\rho}(\xi, \xi'; \beta) = \frac{1}{Z(\beta)}\rho(\xi, \xi'; \beta), \qquad \int \rho(\xi, \xi; \beta)d\xi = Z(\beta), \qquad (4.3)$$

where $Z(\beta)$ is the partition function (statistical sum or statistical integral), the last two names depending generally on the discrete or continuous character of the energy of examined quantum system.

Generally, the canonical density matrix $\rho(\xi, \xi'; \beta)$ can be obtained in two ways:

- *directly*, if we know the energy eigenvalues $E_{n_E}$ of the system's Hamiltonian $H$, then we can use the general definition:

$$\rho(\xi, \xi'; \beta) = \sum_{n_E=1}^{\infty} e^{-\beta E_{n_E}} \Psi_{n_E}(\xi)\Psi_{n_E}^*(\xi'). \qquad (4.4)$$

- *indirectly*, if we do not know the energy eigenvalues $E_{n_E}$, as well as the energy eigenfunctions $\Psi_{n_E}(\xi)$, then we must try to solve the Bloch equation for the canonical density matrix [Feynman, 1972]:

$$-\frac{\partial}{\partial \beta}\rho(\xi, \xi'; \beta) = H(\xi)\rho(\xi, \xi'; \beta) \tag{4.5}$$

with the initial condition:

$$\lim_{\beta \to 0}\rho(\xi, \xi'; \beta) \equiv \rho(\xi, \xi'; 0) = \delta(\xi - \xi') . \tag{4.6}$$

It is true that the Bloch equation (like the Schrödinger equation), even in one dimension, admits precious analytic solutions only for a few quantum systems. Fortunately, for the infinite (continuous and discrete) quantum well this is possible.

In the continuous case, in the *coordinate representation* $x$, the canonical unnormalized density matrix is defined as

$$\rho(x, x'; \beta) = \sum_{n_E=1}^{\infty} e^{-\beta E_{n_E}^{(c)}} \Psi_{n_E}(x) \Psi^*_{n_E}(x') . \tag{4.7}$$

Because, in this case, the density matrix is a real function, it is symmetric with respect to two coordinates: $\rho(x, x'; \beta) = \rho(x', x; \beta)$.

Similarly, for the discrete space, assuming that $x \to x_n = an$ and $x' \to x_n' = an'$ their counterpart will be:

$$\rho(n, n'; \beta) = \sum_{n_E=1}^{\infty} e^{-\beta E_{n_E}^{(d)}} \Psi_{n_E}(n) \Psi^*_{n_E}(n') , \tag{4.8}$$

With $Z^{(c)}(\beta)$ and $Z^{(d)}(\beta)$ we will denote the continuous, respectively discrete statistical sums, defined by normalizing the density matrix to unity:

$$Z^{(c)}(\beta) = \int_0^L \rho(x, x; \beta)dx = \sum_{n_E=1}^{\infty} e^{-\beta E_{n_E}^{(c)}} = \sum_{n_E=1}^{\infty}\left(e^{-\beta \frac{\hbar^2 \pi^2}{2m^* L^2}}\right)^{n_E^2} , \tag{4.9}$$

$$Z^{(d)}(\beta) = a\frac{\Delta^{-1}}{\Delta n}[\rho(n, n; \beta)]_0^N = \sum_{n_E=1}^{\infty} e^{-\beta E_{n_E}^{(d)}} = \sum_{n_E=1}^{\infty} e^{-\beta \frac{\hbar^2}{2m^* a^2}\sin^2\left(\frac{\pi}{N}n_E\right)} . \tag{4.10}$$

The partition function is a quantity of maximal informational importance, because all thermodynamic functions can be expressed through the partition function. But, in the case of free particle in an infinite quantum well (or, precisely, for such a "quantum ideal gas"), due to their mathematical structure, these quantities cannot be expressed analytically and must be calculated numerically or adopted some approximations. In the *Appendix*, we have drawn some considerations regarding the partition function.

However, for the continuous case, the partition function can be evaluated also analytically. So, as an example, let us consider a free electron (with mass $m^* = 9.1 \cdot 10^{-31}$ kg), in an infinite quantum well with width $L = 100 \, \text{Å}$ and equilibrium temperature $T = 300$ K. With $\hbar = 1.054 \cdot 10^{-34}$ J·s and $k_B = 1.38 \cdot 10^{-23}$ J·K$^{-1}$, the exponential $\exp(-\beta E_{n_E}^{(d)})$ into the sum becomes extremely small for any $n_E$, so that the energy levels are very close. Consequently, we can replace the sum with the integral by replacing $y = l n_E$, where we take into account the dimensional aspects:

$$\sum_{n_E=1}^{\infty} \ldots \to \frac{1}{l} \int_0^{\infty} dy \ldots \tag{4.11}$$

In this manner, we obtain:

$$Z^{(c)}(\beta) = \lim_{\substack{N \to \infty \\ a \to 0 \\ aN=L=\text{const}}} Z^{(d)}(\beta) \equiv \lim_{cQM} Z^{(d)}(\beta) = \sum_{n_E=1}^{\infty} \left( e^{-\beta \frac{\hbar^2 \pi^2}{2m^* L^2}} \right)^{n_E^2} \to$$

$$\to \frac{1}{l} \int_0^{\infty} e^{-\beta \frac{\hbar^2 \pi^2}{2m^* L^2} \frac{1}{l^2} y^2} dy = \frac{1}{l} \frac{1}{2} \sqrt{\frac{\pi}{\beta \frac{\hbar^2 \pi^2}{2m^* L^2} \frac{1}{l^2}}} = L \sqrt{\frac{m^*}{2\pi \beta \hbar^2}} \tag{4.12}$$

The last result is the same as in [Feynman, 1972].

Let us solve the Bloch equation for a free particle of mass $m^*$ in an infinite quantum well. In coordinate $x$ - representation for the continuous case this equation reads:

$$-\frac{\partial}{\partial \beta} \rho(x, x'; \beta) = -\frac{\hbar^2}{2m^*} \frac{\partial^2}{\partial x^2} \rho(x, x'; \beta) \tag{4.13}$$

with the initial condition $\rho(x, x'; 0) = \delta(x - x')$.

By passing to the discrete space $x \to x_n = an$ this equation becomes

$$-\frac{\partial}{\partial \beta}\rho(n, n'; \beta) = -\frac{\hbar^2}{2m^*a^2}\frac{\Delta^2}{\Delta n^2}\rho(n, n'; \beta) \quad (4.14)$$

which is a differential-difference equation.

We try to solve this equation by using a new less dimensional thermal variable: $f = \beta\frac{\hbar^2}{2m^*a^2}$. So, we have:

$$\frac{\partial}{\partial f}\rho(n, n'; f) = \frac{\Delta^2}{\Delta n^2}\rho(n, n'; f) \quad (4.15)$$

or, equivalently

$$\frac{\partial}{\partial f}\rho(n, n'; f) = \frac{1}{4}[\rho(n+2, n'; f) - 2\rho(n, n'; f) + \rho(n-2, n'; f)] \quad (4.16)$$

which is a differential-difference equation with partial derivative.

We will try to solve this equation by the separation variable method (similarly as we did for the Bloch equation for the Morse potential, see [Tošić, 2000]), i.e.:

$$\rho(n, n'; f) = \sum_{n_E=1}^{\infty} A_{n_E}(n) B_{n_E}(n'; f) \quad . \quad (4.17)$$

By substituting it into the Bloch equation, we obtain:

$$\sum_{n_E=1}^{\infty} A_{n_E}(n)\frac{\partial}{\partial f}B_{n_E}(n'; f) = \frac{1}{4}\sum_{n_E=1}^{\infty}[A_{n_E}(n+2) - 2A_{n_E}(n) + A_{n_E}(n-2)]B_{n_E}(n'; f) \quad (4.18)$$

which means that

$$\frac{1}{B_{n_E}(n'; f)}\frac{\partial}{\partial f}B_{n_E}(n'; f) = \frac{1}{4}\frac{A_{n_E}(n+2) - 2A_{n_E}(n) + A_{n_E}(n-2)}{A_{n_E}(n)} = -p_{n_E} \quad (4.19)$$

where $p_{n_E}$ is a positive constant which will be determined later.

Now we have to solve the differential equation:

$$\frac{1}{B_{n_E}(n'; f)}\frac{\partial}{\partial f}B_{n_E}(n'; f) = -p_{n_E} \quad (4.20)$$

$$\frac{\partial}{\partial f}B_{n_E}(n'; f) + 0\frac{\partial}{\partial n'}B_{n_E}(n'; f) = -p_{n_E}B_{n_E}(n'; f), \quad (4.21)$$

meaning that we should have:

$$\frac{df}{1} = \frac{dn'}{0} \quad ; \qquad \frac{df}{1} = -\frac{1}{p_{n_E}} \frac{dB_{n_E}(n';f)}{B_{n_E}(n';f)} \qquad (4.22)$$

$$\frac{dB_{n_E}(n';f)}{B_{n_E}(n';f)} = -p_{n_E} df \quad ; \qquad B_{n_E}(n';f) = B_{n_E}(n';0)\, e^{-f\, p_{n_E}} \quad , \qquad (4.23)$$

where, at the moment, $B_{n_E}(n';0)$ is an arbitrary function, constant with respect to variable $f$, which will be determined later, from the symmetry condition of the density matrix (which, in our case is also a real function).

The second equation

$$A_{n_E}(n+2) - 2(1 - 2p_{n_E})A_{n_E}(n) + A_{n_E}(n-2) = 0 \qquad (4.24)$$

is a difference equation like the Schrödinger equation for a free particle in the infinite quantum well. For this reason, we can perform the identification, as well as the quantization condition:

$$p_{n_E} = \frac{2m^* E_{n_E}^{(d)}}{\hbar^2} a^2 = \sin^2 \frac{\pi}{N} n_E \ . \qquad (4.25)$$

Also, from the previous resolution of the Schrödinger equation, we take the solution:

$$A_{n_E}(n) = \tilde{C} \sin\left(\pi \frac{n_E}{N} n\right) . \qquad (4.26)$$

Consequently, the density matrix becomes

$$\rho(n, n'; f) = \tilde{C} \sum_{n_E=1}^{\infty} e^{-f\, p_{n_E}} \sin\left(\pi \frac{n_E}{N} n\right) B_{n_E}(n';0) \ . \qquad (4.27)$$

Because the eigenfunctions $\Psi_{n_E}(n)$ are real functions, the density matrix is symmetric, so the function $B_{n_E}(n';0)$ must be:

$$B_{n_E}(n';0) = \tilde{C} \sin\left(\pi \frac{n_E}{N} n'\right) \qquad (4.28)$$

If we return to the old notations

$$f\, p_{n_E} = \beta E_{n_E}^{(d)} \qquad (4.29)$$

the density matrix becomes

$$\rho(n, n'; \beta) = \tilde{C}^2 \sum_{n_E=1}^{\infty} e^{-\beta E_{n_E}^{(d)}} \sin\left(\frac{\pi n_E}{N} n\right) \sin\left(\frac{\pi n_E}{N} n'\right). \tag{4.30}$$

Having in mind the expression of the discrete eigenfunctions $\Psi_{n_E}(n)$, it follows that we have $\tilde{C} = C_2 = \sqrt{\frac{2}{L}}$ and so, the canonical density matrix becomes:

$$\rho(n, n'; \beta) = \frac{2}{L} \sum_{n_E=1}^{\infty} e^{-\beta E_{n_E}^{(d)}} \sin\left(\frac{\pi n_E}{N} n\right) \sin\left(\frac{\pi n_E}{N} n'\right), \tag{4.31}$$

i.e. exact the same mathematical structure as the one obtained from the general definition of the discrete canonical density matrix.

Like the density matrix for the continuous case, their counterpart for the discrete one can be also normalized to unity and consequently, the discrete integral from the canonical density matrix is just the partition function $Z^{(d)}(\beta)$:

$$a \frac{\Delta^{-1}}{\Delta n} \rho(n, n; \beta) \Big|_0^N = \frac{2}{L} \sum_{n_E=1}^{\infty} e^{-\beta E_{n_E}^{(d)}} a \frac{\Delta^{-1}}{\Delta n} \left[\sin^2\left(\frac{\pi n_E}{N} n\right)\right]_0^N = Z^{(d)}(\beta). \tag{4.32}$$

The discrete integral in the above relation was previously calculated (see Sec.3) and their value is $N/2$. Consequently, we obtain the correct expression for the partition function $Z^{(d)}(\beta)$.

In this manner, *the normalized* discrete density matrix becomes

$$\tilde{\rho}(n, n'; \beta) \equiv \frac{1}{Z^{(d)}(\beta)} \rho(n, n'; \beta) = \frac{2}{L} \frac{1}{Z^{(d)}(\beta)} \sum_{n_E=1}^{\infty} e^{-\beta E_{n_E}^{(d)}} \sin\left(\frac{\pi n_E}{N} n\right) \sin\left(\frac{\pi n_E}{N} n'\right). \tag{4.33}$$

On the other hand, the canonical density matrix (i.e. the solution of the Bloch equation) can fulfill the initial condition with respect to the variable $\beta$. So, we have:

$$\lim_{\beta \to 0} \rho(n, n'; \beta) = \frac{2}{L} \sum_{n_E=1}^{\infty} \sin\left(\pi \frac{a n}{L} n_E\right) \sin\left(\pi \frac{a n'}{L} n_E\right) = \delta(a n - a n') = \frac{1}{a} \delta(n - n') \tag{4.34}$$

where we have used the Fourier series representation for the Dirac delta function [Delta]:

$$\delta(x - x') = \frac{2}{L} \sum_{k=1}^{\infty} \sin\left(\pi \frac{x}{L} k\right) \sin\left(\pi \frac{x'}{L} k\right), \quad L > 0. \tag{4.35}$$

Let us apply the continuous limit to the canonical discrete density matrix $\rho(n, n'; \beta)$. We obtain:

$$\rho(x, x'; \beta) = \lim_{\substack{N \to \infty \\ a \to 0 \\ aN=L=\text{const}}} \rho(n, n'; \beta) \equiv \lim_{cQM} \rho(n, n'; \beta) =$$

$$= \frac{2}{L} \sum_{n_E=1}^{\infty} \lim_{cQM} \left[ e^{-\beta E_{n_E}^{(d)}} \right] \sin\left(\frac{\pi n_E}{L} x\right) \sin\left(\frac{\pi n_E}{L} x'\right) = \quad (4.36)$$

$$= \frac{2}{L} \sum_{n_E=1}^{\infty} e^{-\tilde{\beta} n_E^2} \sin(\alpha n_E) \sin(\alpha' n_E)$$

where $\tilde{\beta} \equiv \beta \frac{\hbar^2 \pi^2}{2m^* L^2}$, $\alpha \equiv \pi \frac{x}{L}$ and $\alpha' \equiv \pi \frac{x'}{L}$.

As we have seen earlier, the quantity $\tilde{\beta} \equiv \beta \frac{\hbar^2 \pi^2}{2m^* L^2}$ is very small and we can replace the sum with respect to $n_E$ by integral with respect to a continuous variable, say $y = l n_E$ as earlier. If we develop the sinus products using the Euler's formulae

$$\sin(\alpha n_E)\sin(\alpha' n_E) = \frac{1}{4}\left[ e^{i(\alpha-\alpha')n_E} + e^{-i(\alpha-\alpha')n_E} - e^{i(\alpha+\alpha')n_E} - e^{-i(\alpha+\alpha')n_E} \right] \quad (4.37)$$

then we have to calculate four integrals:

$$\lim_{cQM} \rho(n, n'; \beta) = \frac{1}{2L}\left[ I_1^- + \left(I_1^-\right)^* - I_2^+ - \left(I_2^+\right)^* \right] \quad (4.38)$$

where:

$$I_1^- \equiv \frac{1}{l}\int_0^\infty e^{-\frac{\tilde{\beta}}{l^2} y^2 + i\frac{(\alpha-\alpha')}{l} y} dy = \frac{1}{2}\sqrt{\frac{\pi}{\tilde{\beta}}} e^{-\frac{(\alpha-\alpha')^2}{4\tilde{\beta}}} = L\sqrt{\frac{m^*}{2\pi\beta\hbar^2}} e^{-\frac{m^*}{2\beta\hbar^2}(x-x')^2} \quad (4.39)$$

$$I_2^+ \equiv \frac{1}{l}\int_0^\infty e^{-\frac{\tilde{\beta}}{l^2} y^2 + i\frac{(\alpha+\alpha')}{l} y} dy = \frac{1}{2}\sqrt{\frac{\pi}{\tilde{\beta}}} e^{-\frac{(\alpha+\alpha')^2}{4\tilde{\beta}}} = L\sqrt{\frac{m^*}{2\pi\beta\hbar^2}} e^{-\frac{m^*}{2\beta\hbar^2}(x+x')^2}. \quad (4.40)$$

These integrals are of the following kind (see, 3.323/2 of Ref. [Gradshteyn, 2007]):

$$\int_0^\infty e^{-p^2\left(y \pm i\frac{q}{2p^2}\right)^2} dy = \frac{1}{2}\sqrt{\frac{\pi}{p^2}} e^{-\frac{q^2}{4p^2}}. \quad (4.41)$$

Note that the integrals do not depend on the sign in front of $(\alpha \pm \alpha')$ and thus, $I_1^- = \left(I_1^-\right)^*$ and $I_2^+ = \left(I_2^+\right)^*$. Apart from this, the last two integrals, $I_2^+$ and $\left(I_2^+\right)$, which finally contain the sum $(x+x')^2$, do not lead to the Dirac delta function at the limit $\beta \to 0$, so they have no significance.

Finally, we obtain the same results as in [Feynman, 1972]:

$$\rho(x, x'; \beta) = \sqrt{\frac{m^*}{2\pi\beta\hbar^2}} e^{-\frac{m^*}{2\beta\hbar^2}(x-x')^2}. \tag{4.42}$$

By normalizing this density matrix, we obtain the same expression of the partition function as the one obtained by applying the continuous limit (see, Eq. (4.12)) to the discrete partition function $Z^{(d)}(\beta)$:

$$Z^{(c)}(\beta) = \int_0^L \rho(x, x; \beta) dx = L\sqrt{\frac{m^*}{2\pi\beta\hbar^2}} = e^{-\beta F}, \tag{4.43}$$

where $F$ is the free energy of the quantum gas in the infinite quantum well [Feynman, 1972].

This is another proof of the correctness of the calculations made for the discrete case of the free quantum particle in the infinite quantum well.

The normalized density matrix then becomes:

$$\tilde{\rho}(x, x'; \beta) = \frac{1}{L} e^{-\frac{m^*}{2\beta\hbar^2}(x-x')^2}. \tag{4.44}$$

Let us we calculate now the thermal expectation of the Hamiltonian $H = -\frac{\hbar^2}{2m^*}\frac{\partial^2}{\partial x^2}$ of free particle trapped in the infinite square quantum well, by using the canonical density matrix $\rho(x, x'; \beta)$.

In the cQM this is given by the expression [Feynman, 1972]:

$$<H>^{(c)} = -\frac{\hbar^2}{2m^*}\left\langle \frac{\partial^2}{\partial x^2} \right\rangle = -\frac{\hbar^2}{2m^*}\frac{1}{Z_c(\beta)}\int_0^L dx \left[\frac{\partial^2}{\partial x'^2}\rho(x,x';\beta)\right]_{x'=x} \tag{4.45}$$

where we must perform the following succession of operations: first we carry out the differentiation with respect to the variable $x'$, then we replace $x'$ with $x$ and finally we integrate with respect to the variable $x$.

Then, in the dQM, the corresponding expression will be

$$<H>^{(d)} = -\frac{\hbar^2}{2m^*}\left\langle \frac{1}{a^2}\frac{\Delta^2}{\Delta n^2}\right\rangle = -\frac{\hbar^2}{2m^*}\frac{1}{Z_c(\beta)}\left\{a\frac{\Delta^{-1}}{\Delta n}\left[\frac{1}{a^2}\frac{\Delta^2}{\Delta n'^2}\rho(n,n';\beta)\right]_{n'=n}\right\}\Bigg|_0^N \quad (4.46)$$

with the same succession of operations as earlier.

The discrete differentiation is easy to be performed if we use the Bloch equation (4.14), i.e. if we observe the operator equality which are applied to the function $\rho(n,n';\beta)$:

$$\frac{\partial}{\partial\beta} = \frac{\hbar^2}{2m^*a^2}\frac{\Delta^2}{\Delta n^2} \quad (4.47)$$

Then the expectation value becomes

$$<H>^{(d)} = -\frac{\hbar^2}{2m^*}\left\langle\frac{1}{a^2}\frac{\Delta^2}{\Delta n^2}\right\rangle = -\left\langle\frac{\partial}{\partial\beta}\right\rangle = -\frac{1}{Z_d(\beta)}\left\{a\frac{\Delta^{-1}}{\Delta n}\left[\frac{\partial}{\partial\beta}\rho(n,n';\beta)\right]_{n'=n}\right\}\Bigg|_0^N =$$

$$= -\frac{1}{Z_d(\beta)}\frac{2}{N}\frac{\partial}{\partial\beta}\sum_{n_E=1}^{\infty}e^{-\beta E_{n_E}^{(d)}}\left\{\frac{\Delta^{-1}}{\Delta n}\left[\sin^2(n\varphi)\right]\right\}\Bigg|_0^N = -\frac{\partial}{\partial\beta}\ln Z_d(\beta) \quad (4.48)$$

where we have used Eq. (2.25), the notation $\varphi = \frac{n_E}{N}\pi$ and the equality $\sin(2\pi n_E) = 0$.

In this manner, for dQM we have obtained the correct results, like for cQM.

All the obtained results for the case of quantum free particles and a quantum non interacting particle (ideal) gas trapped in a discrete infinite quantum well are summarized in the *Table 1*, from *Appendix* 1, in comparison with the same continuous case.

### 5. Concluding remarks

In the present paper, we presented a manner of treating the infinite square quantum well in discrete quantum mechanics using a central difference second-derivatives approach. The results obtained by discretization of both the Schrödinger and the Bloch equations could be of certain interest in a different context, especially in the condensed matter physics.

The aim of our paper was to solve the Bloch equation for quantum non interacting particles (the „quantum ideal gas") trapped in a square quantum well of discretized space of the width $L$. For this purpose, at the beginning, we recalled some (earlier known) methods and results regarding the solving of the Schrödinger equation of a quantum particle trapped in the square infinite well, in the discrete space. The depth of the quantum well is considered infinite $U \to \infty$ so, inside the well the potential is zero.

If we compare the two approaches (discrete and continuous) for the purpose of studying the behavior of the free particle (or the "quantum ideal gas" of free particles) in the infinite quantum well of the same width, we can observe some fundamental differences. The paper lead to the conclusion that, for the problems which include the resolution of the Schrödinger equation (respectively, the Bloch equation) in the discrete space for a particle (respectively, non interacting particle gas) trapped in an infinite square well, the most useful form of the discrete derivatives (between forward, backward and central variants) is the central difference formula, because this approach is consistent with the oscillatory behaviour of the wave functions of the examined problem. The main difference refers to the energy spectrum and, consequently, to the band structure. If in the continuous case the energy has parabolic dependence with respect to the main or principal quantum number $n_E$, in the discrete case this dependence is more complicated, being achieved by means of the square sinus function $\sin^2\left(\pi \frac{n_E}{N}\right)$.

By applying the continuous quantum mechanical limit ($\lim_{cQM}$) to the physical quantities or formulae regarding the discrete case $\mathcal{A}^{(d)}$, we lead to the corresponding quantities or formulae for the continuous case $\mathcal{A}^{(c)}$:

$$\lim_{\substack{N \to \infty \\ a \to 0 \\ aN=L=\text{const}}} \mathcal{A}^{(d)} \equiv \lim_{cQM} \mathcal{A}^{(d)} = \mathcal{A}^{(c)}$$

As we pointed out, the main purpose of the paper was to solve the discretized Bloch equation for the free particle in the infinite quantum well which is, in our opinion, a completely new approach which had not appeared yet in the scientific literature.

The discretized approach is appropriate for many problems e.g. in condensed matter physics, because in this field many physical systems are achieved from crystalyne semiconductors wherein the spacing between atoms is finite.

On the other hand, the discretization approach can be used successfully in different nanostructures including quantum wells, wires and dots, i.e. physical systems containing a finite number of atoms.

In conclusion, together with other works examining some aspects of the quantum infinite square well, for example tight-binding models [Boykin, 2004], [Boykin, 2005], coherent states [Garcia de Leon, 2008], [Popov, 2014], or those using Lanczos procedure [Groenenboom, 1990], we believe that our paper will be a small step forward and will contribute to a better understanding of the phenomena involving quantum infinite square well (especially as regarded by the Bloch equation).

**Appendix 1**

**Table 1. Discrete versus continuous approach for the infinite quantum well**

| | Discrete approach: *Infinite discrete quantum well* $\mathcal{A}^{(d)}$ | Continuous approach: *Infinite continuous quantum well* $\mathcal{A}^{(c)}$ |
|---|---|---|
| Variable | $x_n = a\,n$ - „discrete points" | $x$ - coordinate |
| Range | $n = 0, 1, 2, \ldots, N$; $L = N a$; $a > 0$ | $x \in [0, L]$; $L > 0$ |
| Case connection | $\lim\limits_{\substack{N \to \infty \\ a \to 0 \\ aN=L=\text{constant}}} A^{(d)} \equiv \lim\limits_{cQM} A^{(d)} = A^{(c)}$ ||
| Principal quantum number | $n_E = 1, 2, \ldots, \infty$ | $n_E = 1, 2, \ldots, \infty$ |
| Eigenenergy | $E_{n_E}^{(d)} = \dfrac{\hbar^2}{2m^* a^2} \sin^2\left(\pi \dfrac{n_E}{N}\right)$ | $E_{n_E}^{(c)} = \dfrac{\hbar^2 \pi^2}{2m^* L^2} n_E^2$ |
| Eigenfunction | $\Psi_{n_E}(n) = \sqrt{\dfrac{2}{L}} \sin\left(\pi \dfrac{n_E}{N} n\right)$ | $\Psi_{n_E}(x) = \sqrt{\dfrac{2}{L}} \sin\left(\pi \dfrac{n_E}{L} x\right)$ |
| Schrödinger equation | $\dfrac{\Delta^2}{\Delta n^2} \Psi(n) + \dfrac{2m^* E}{\hbar^2} a^2 \Psi(n) = 0$ | $\dfrac{d^2}{dx^2} \Psi(x) + \dfrac{2m^* E}{\hbar^2} \Psi(x) = 0$ |
| Bloch equation | $-\dfrac{\partial}{\partial\beta}\rho(n,n';\beta)=$ $=-\dfrac{\hbar^2}{2m^*a^2}\dfrac{\Delta^2}{\Delta n^2}\rho(n,n';\beta)$ | $-\dfrac{\partial}{\partial\beta}\rho(x,x';\beta)=$ $=-\dfrac{\hbar^2}{2m^*}\dfrac{\partial^2}{\partial x^2}\rho(x,x';\beta)$ |
| Initial condition | $\lim_{\beta\to 0}\rho(n,n';\beta)\equiv\rho(n,n';0)=$ $=\dfrac{1}{a}\delta(n-n')$ | $\lim_{\beta\to 0}\rho(x,x';\beta)\equiv\rho(x,x';0)=$ $=\delta(x-x')$ |
| Canonical density matrix | $\rho(n,n';\beta)=$ $=\dfrac{2}{L}\sum_{n_E=1}^{\infty}e^{-\beta E_{n_E}^{(d)}}\sin\!\left(\dfrac{\pi n_E}{N}n\right)\sin\!\left(\dfrac{\pi n_E}{N}n'\right)$ | $\rho(x,x';\beta)=\sqrt{\dfrac{m^*}{2\pi\beta\hbar^2}}\,e^{-\dfrac{m^*}{2\beta\hbar^2}(x-x')^2}$ |
| Partition function | $Z_d(\beta)=\sum_{n_E=1}^{\infty}e^{-\beta\dfrac{\hbar^2}{2m^*a^2}\sin^2\!\left(\dfrac{\pi}{N}n_E\right)}$ | $Z_c(\beta)=\sum_{n_E=1}^{\infty}\left(e^{-\beta\dfrac{\hbar^2}{2m^*a^2N^2}\pi^2}\right)^{n_E^2}=$ $=L\sqrt{\dfrac{m^*}{2\pi\beta\hbar^2}}$ |
| Normalized density matrix | $\tilde\rho(n,n';\beta)=$ $=\dfrac{2}{L}\dfrac{1}{Z^{(d)}(\beta)}\sum_{n_E=1}^{\infty}e^{-\beta E_{n_E}^{(d)}}\sin\!\left(\dfrac{\pi n_E}{N}n\right)\sin\!\left(\dfrac{\pi n_E}{N}n'\right)$ | $\tilde\rho(x,x';\beta)=\dfrac{1}{L}e^{-\dfrac{m^*}{2\beta\hbar^2}(x-x')^2}.$ |

**Appendix 2**

The evaluation in an analytical manner of the partition function of the canonical partition function, both for the discrete and also for the continuous case would be an extremely difficult task [Wolf, 1990].

Due to the fact that $0<p_{n_E}=\sin^2\!\left(\pi\dfrac{n_E}{N}\right)\le 1$, as well as $\dfrac{\hbar^2}{2m^*a^2}$ is very small, the energy levels are very close. Therefore, we try to express the canonical partition function of the discrete case in an approximate manner, by reteining only two terms of the sum. This is motivated by the fact that in the low temperature limit, the exponent is exceedingly small, so it may consider that only the first two terms make a most important contribution (see the numerical example in the Section 4). We denote this approximation with $Z_d^{(2)}(\beta)$:

$$Z_d^{(2)}(\beta)=e^{-\beta E_1^{(d)}}+e^{-\beta E_2^{(d)}} \quad . \tag{A.1}$$

This approximation correspond to a two-level system (e.g. for a particle with spin ½ placed in a magnetic field $B$ along the $z$ direction, when $E_{1,2}^{(d)} = \pm \mu_B B$ ).

Let us use this expression in order to calculate the molar heat capacity at the constant volume $C_V^{(2)}$, defined as:

$$C_V^{(2)} = \frac{1}{\nu} \frac{\partial}{\partial T} \left( N_{tot} <H>^{(d)} \right) = R\beta^2 \frac{\partial^2}{\partial \beta^2} \ln Z_d^{(2)}. \tag{A.2}$$

After some simple algebraic calculations we arrived at an interesting final expression:

$$\frac{C_V^{(2)}}{R} = \left[ \frac{\beta \frac{E_1^{(d)} - E_2^{(d)}}{2}}{\cosh\left( \beta \frac{E_1^{(d)} - E_2^{(d)}}{2} \right)} \right]^2 . \tag{A.3}$$

Using the notation

$$\Delta E_{1,2}^{(d)} \equiv E_1^{(d)} - E_2^{(d)} = \frac{\hbar^2}{2m^* a^2} \left[ \sin^2\left(\frac{\pi}{N}\right) - \sin^2\left(2\frac{\pi}{N}\right) \right] \tag{A.4}$$

we can introduce a quantity analogue of the Debye characteristic temperature $\Theta_{1,2}^{(d)}$:

$$\frac{\Theta_{1,2}^{(d)}}{T} \equiv \frac{1}{2} \frac{\Delta E_{1,2}^{(d)}}{k_B} \tag{A.5}$$

so that the molar heat capacity at the constant volume $C_V^{(2)}$ becomes

$$\frac{C_V^{(2)}}{R} = \left[ \frac{\frac{\Theta_{1,2}^{(d)}}{T}}{\cosh\left( \frac{\Theta_{1,2}^{(d)}}{T} \right)} \right]^2 \tag{A.6}$$

As a function of temperature $T$, for low temperatures (up to 300 K), this expression for the caloric capacity at constant volume has the same aspect as the vibrational heat capacity in the Einstein's model or thermal capacity of spin ½ paramagnet.

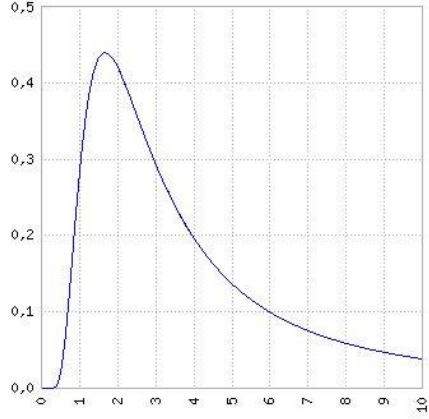

**Fig. 1** *The graphical behavior of* $\dfrac{C_V^{(2)}}{R}$ *as a function of* $T$, *according to Eq.(A.3).*

As a matter of fact we indicate that the continuous canonical partition function $Z^{(c)}(\beta)$ (4.9) can be evaluated, up to an additional constant, by the help of the Jakobi theta function which is an integral functions defined as follows [Theta]:

$$\vartheta_3(z,\mu) = \sum_{n_E=-\infty}^{\infty} \exp\{-i 2 z n_E - \mu n_E^2\} = \sqrt{\frac{\pi}{\mu}} \sum_{n_E=-\infty}^{\infty} \exp\left\{-\frac{(z-\pi n_E)^2}{\mu}\right\}. \qquad (A.5)$$

For $z=0$ we have $\vartheta_3(0,\mu) \equiv \vartheta_3(\mu)$, i.e. [Munoz Villegas, 2003]:

$$\vartheta_3(\mu) = \sum_{n_E=-\infty}^{\infty} \exp\{-\mu n_E^2\} = \sqrt{\frac{\pi}{\mu}} \sum_{n_E=-\infty}^{\infty} \exp\left\{-\frac{\pi^2}{\mu} n_E^2\right\} = \sqrt{\frac{\pi}{\mu}} + O\left(e^{-\frac{\pi^2}{\mu}}\right). \qquad (A.6)$$

For $\mu \ll 1$ we can neglect the correction terms to order $e^{-\frac{\pi^2}{\mu}}$ and so, we have

$$\vartheta_3(\mu) = \sqrt{\frac{\pi}{\mu}}. \qquad (A.7)$$

Consequently, by identifying $\mu = \beta \dfrac{\hbar^2 \pi^2}{2m^* L^2}$, the continuous partition function $Z^{(c)}(\beta)$ becomes:

$$Z^{(c)}(\beta) = \sum_{n_E=1}^{\infty} e^{-\beta \frac{\hbar^2 \pi^2}{2m^* L^2} n_E^2} = \frac{1}{2}\left(\sum_{n_E=-\infty}^{\infty} e^{-\beta \frac{\hbar^2 \pi^2}{2m^* L^2} n_E^2} - 1\right) = \frac{1}{2}\left[\vartheta_3\left(\beta \frac{\hbar^2 \pi^2}{2m^* L^2}\right) - 1\right] = L\sqrt{\frac{m^*}{2\pi \hbar \beta}} - \frac{1}{2}.$$

(A.8)

Thus, with an accuracy up to a constant term, which does not contribute to the heat capacity value, we obtained in another way Eq. (4.12).

**Acknowledgments**